\documentclass[a4paper,11pt]{article}
\usepackage{epsfig,amssymb,amstext}

\newcommand{\rem}[1]{}
\newcommand{\sem}[1]{#1}
\newcommand{\secref}{\section*{References}}
 
 \newcommand{\ack}{\section*{Acknowledgement}}
 \newcommand{\address}[1]{}
 \newcommand{\ftnote}[2]{}
 \newcommand{\jl}[1]{}
 \newcommand{\pacs}[1]{\vspace{1ex} \noindent PACS: #1}
 \renewcommand{\secref}{}

\newcommand{\tr}{{\rm tr}} 

\newcommand{\Goodbreak}{\goodbreak}

\newcommand{\pafra}[2]{\frac{\partial #1}{\partial #2}}
\newcommand{\MOf}{{\bf M}^{(4)}}

\newcommand{\cau}{\mu} 
\newcommand{\cS}{{\bf S}_{\rm i}}
\newcommand{\cC}{{\bf C}_{\rm o}}
\newcommand{\Oms}{{\bf \Omega}}
\newcommand{\diag}{{\rm diag}}

\newcommand{\Ap}{{\cal A}}              
\newcommand{\Angle}{\psi}     
\newcommand{\Bmag}{B}           
\newcommand{\A}{{\bf A}}        
\newcommand{\const}{\mbox{const}}       
\newcommand{\C}{{\bf C}}        
\newcommand{\CK}{{\bf R}}       
\newcommand{\CKe}{R}            
\newcommand{\D}{{\bf D}}        
\newcommand{\Eh}{{\cal E}_h}    
\newcommand{\Epn}{E}            
\newcommand{\F}{F}              
\newcommand{\g}{{\bf g}}        
\newcommand{\I}{I}              
\newcommand{\J}{{\bf J}}        
\newcommand{\K}{{\bf K}}        
\renewcommand{\L}{{\bf L}}      
\newcommand{\M}{{\bf M}}        
\newcommand{\mojac}{Jacobian}   
\newcommand{\n}{n}              
\newcommand{\N}{{\bf n}}        
\newcommand{\p}{{p}}    
\renewcommand{\P}{{\bf p}}      
\newcommand{\DQ}[2]{\pafra{\XX_{#1}}{\XX_{#2}}}
\newcommand{\Qh}{Q_h}           
\newcommand{\refmap}{\rho}      
\renewcommand{\r}{{\bf q}}        
\newcommand{\rr}{r}             
\newcommand{\qrx}{x}
\newcommand{\qry}{y}
\newcommand{\R}{{\bf r}}        
\newcommand{\Real}{{\mathbb R}}    
\newcommand{\GR}{R}             
\newcommand{\isig}{\sigma^{-1}} 
\renewcommand{\S}{{\bf S}}      
\newcommand{\Sh}{\Sigma_h}      
\newcommand{\T}{{\bf t}}        
\newcommand{\trt}{{\tr}^{(2)}}       
\newcommand{\Trf}{{\tr}^{(4)}}       
\renewcommand{\u}{{\bf u}}      
\newcommand{\nut}{\gamma}       
\renewcommand{\v}{{\bf v}}      
\newcommand{\W}{{\bf W}}        
\newcommand{\XX}{\Xi}   
\newcommand{\XP}{X}     

\newcommand{\ps}{\partial_s}
\newcommand{\pp}{\partial_\p}

\newcommand{\pr}{\partial_q}            
\newcommand{\ph}{\partial_h}

\newcommand{\SP}[2]{\langle #1, #2\rangle}
\newcommand{\vecII}[2]{\left(\begin{array}{c} #1 \\ #2 \end{array}\right)}
\newcommand{\matII}[2]{\left(\begin{array}{cc} #1 \\ #2 \end{array}\right)}
\newcommand{\matIIIV}[2]{\left(\begin{array}{cccc} #1 \\ #2 \end{array}\right)}

\renewcommand{\[}{\begin{equation}}
\renewcommand{\]}{\end{equation}}

\renewcommand{\|}{\,|\,}
\newcommand{\Id}{{\bf 1}}       
\newcommand{\Nu}{{\bf 0}}       

\newcommand{\weg}[1]{}

\newcommand{\fig}{Fig.}
\rem{
\newtheorem{platz}{{\bf Fig.}} 
\newcommand{\Capts}[1]{#1}
\newcommand{\FIG}[3]{\vspace*{-4ex}
  \marginpar{ \begin{platz} \label{#1} ~ \end{platz} }\vspace*{4ex}}
\newcommand{\printfigcap}[2]{\rule{0cm}{1cm}{\bf Figure~\ref{#1}:} #2 \\ }
\newcommand{\showfig}[1]{\begin{figure} #1 \caption{H.R.~Dullin: Linear stability in billiards with potential} \end{figure} } 
} 
\newcommand{\Capts}[1]{}
\newcommand{\FIG}[3]{\begin{figure}
  #3 \caption[]{\small #2} \label{#1} \end{figure}}

\begin{document}

\jl{8}

\rem{
\title{Linear stability in billiards with potential}
\author{Holger R. Dullin\ftnote{1}{E-mail address: hdullin@physik.uni-bremen.de}
}
\address{Institut f\"ur Theoretische Physik,
Universit\"at Bremen,
Postfach 330440,
28344 Bremen, Germany 
}
\date{}
}

\sem{ 
\title{Linear stability in billiards with potential}
\author{Holger R. Dullin \\
Institut f\"ur Theoretische Physik \\
Universit\"at Bremen \\
Postfach 330440 \\
28344 Bremen, Germany \\
Email: hdullin@physik.uni-bremen.de
}
\date{}
\maketitle
} 

\begin{abstract}
A general formula for the linearized Poincar\'e map of a billiard
with a potential is derived. The stability of periodic orbits 
is given by the trace of a product of matrices describing the
piecewise free motion between reflections and the contributions
from the reflections alone. For the case without potential this
reduces to well known formulas.
Four billiards with potentials for which the free motion is integrable
are treated as examples: The linear gravitational potential, the
constant magnetic field, the harmonic potential, and a billiard in a
rotating frame of reference, imitating the restricted three body problem.
The linear stability of periodic orbits with period one and two is
analyzed with the help of stability diagrams, showing the essential
parameter dependence of the residue of the periodic orbits for these
examples.

\end{abstract}


\section{Introduction}

In the theory of dynamical systems billiards were introduced 
in the twenties of this century by Birkhoff. His idea was to have
a simple class of models that show the complicated behavior 
of nonintegrable smooth Hamiltonian systems without the need to 
integrate a differential equation \cite{Birkhoff27}: ``But in this
problem the formal side, usually so formidable in dynamics, almost
completely disappears, and only the interesting qualitative questions
need to be considered''.
Birkhoff studied billiards inside smooth convex domains, for which it 
is now known \cite{Lazutkin73} that they can not be ergodic.
In the seventies Sinai proved the ergodicity of the billiard on
a torus with a disk removed \cite{Sinai70}, and subsequently 
Bunimovich, Wojtkowski and others \cite{Bunimovich79,Wojtkowski86}
designed larger classes of ergodic billiards.
Starting with some early papers in the eighties \cite{BenS78,Berry81}
billiards became quite popular in physics.
For investigations of both, the classical dynamics and
the quantum mechanical properties, periodic orbits and their 
properties are of major interest, which are investigated using
the Poincar\'e map.

For billiards the natural Poincar\'e map is from bounce to bounce:
The boundary $\partial \Omega$ of the billiard domain $\Omega \subset \Real^2$ 
is parametrized by the arc length $s$. 
$\partial \Omega$ is assumed to be $C^2$, such that the curvature
is well defined. Reflections at points in the boundary which are 
less smooth need a special treatment.
Starting at some point $\R(s_1) \in \partial \Omega$
all possible orbits leaving $\partial \Omega$ 
can be parametrized by an angle $\beta$ from $-\pi/2$ to $\pi/2$, 
where $-\pi/2$, $0$, $\pi/2$ specify the directions $-\T(s_1)$, 
$\N(s_1)$, $\T(s_1)$ 
where $\T(s_1)$ is the tangent vector to $\partial \Omega$ at $\R(s_1)$ 
and $\N(s_1)$ is the normal vector pointing into $\Omega$. 
We have $\ps \R(s) = \T(s)$ and $\ps \T(s) = \kappa(s) \N(s)$
where $\kappa(s)$ is the signed curvature at $s$, 
negative $\kappa$ indicating a defocusing boundary with respect to
$\Omega$ or, equivalently, the normal $\N$.
If instead of $\beta$ the component $p$ of the velocity 
parallel to $\T$ is used, $\p=\sin\beta$, then $(s,\p)$ are
canonical coordinates on the Poincar\'e surface of section.
The trajectory starting with $\xi_1 = (s_1,\p_1)$ encounters the boundary 
after time $\tau$ at $s_2$ and has a parallel component $p_2$ 
(with respect to $\T(s_2)$) before and after the reflection.
The Poincar\'e map is denoted by $P$ such that $P\xi_1 = \xi_2$.

A periodic orbit of period $n$ is now described by a list of points 
$\xi_i = (s_i,\p_i)$, $i=1,\ldots,n$ such that $P\xi_i = \xi_{i+1}$ 
for $i=1,\ldots,n-1$ and $P\xi_n = \xi_1$, or equivalently
$P^n \xi_1 = \xi_1$.
Finding periodic orbits is a hard problem, even for billiards. 
In this paper we take the point 
of view that the periodic orbits are given to us, and we want to 
find a formula for their stability, i.e.\ for the linearized map $DP$ of $P$.

In the following we will denote a quantity to be 
evaluated at $\xi_i$ or $s_i$ by a lower index $i$, 
e.g.\ $\R(s_i) = \R_i$, $\T(s_i) = \T_i$ or
$DP_{i,i+1}$ for the linearized map of $P$ from $\xi_i$
to $\xi_{i+1}$. Then we obtain for $DP^n$
\[
        DP^n_{n,1} = DP_{n,n-1} \cdots DP_{32} DP_{21}
\]
by the chain rule. For a smooth Hamiltonian system the linearized map 
depends on the solution of the variational differential
equation, which can usually only be obtained by numerical integration.
For an ordinary billiard without potential, 
rather than the infinite amount of information contained in
this solution, we just need three numbers per reflection,
besides the initial and final point: 
the time $\tau$ (or length) between reflections, and the curvatures
$\kappa_i$ and $\kappa_{i+1}$ of $\partial \Omega$ at the initial 
and final points $\R_i$ and $\R_{i+1}$.
The expression of $DP$ in terms of these data is well 
known, see e.g.~\cite{HenWis83,KT91,DRW96}. 
The main point in stability formulas of this type is that they do {\em not}
express everything in terms of the initial point $\xi$, which is of course
possible. Instead, some quantities from the free motion, 
e.g.\ the time of flight $\tau$, directly enter the formula. 
The time $\tau$ is usually given by the solution of a complicated
equation (only for special boundaries can it be obtained from the
roots of a polynomial of degree up to four). Therefore it makes sense
(also numerically) to use $\tau$ in stability formulas, because it
must have been already calculated to obtain $P$.

If we add a simple potential to the billiard, the problem of the
calculation of $DP$ becomes a little harder. To keep the original
idea of Birkhoff it only makes sense to add integrable potentials
for which the solutions of the free motion in the potential can
be written down explicitly. The general formula we obtain does not
depend on the integrability, however in the examples we only treat
integrable potentials. Typical examples are linear potentials
\cite{LeMi86,RSW90,SG92,Chaplygin33,HayVid94,KT91,KL91} 
with free motion along parabolas, 
quadratic potentials \cite{Jacobi1866,KT91} 
with free harmonic motion,
magnetic fields \cite{RobnikBerry85,Berglund94} with free motion on circles
and the fictitious forces of a rotating frame of reference \cite{MBLTJS95}.
The reason for the study of billiards with potential, besides of course
the situation where they model a given physical situation, e.g.\ the
restricted three body problem in \cite{MBLTJS95}, is that the potential
can introduce two typical features of Hamiltonian systems which are
absent in ordinary billiards. The most important is the energy dependence:
As it is nicely illustrated in \cite{HayVid94}, one can observe the typical
evolution from an integrable low energy limit, via an intermediate regime 
with the most complicated motion to an integrable high energy limit.
This scenario is typical for bounded Hamiltonian systems, and
billiards with potential are simple
model systems of this kind.
The second reason is that a magnetic field (or a rotating system)
breaks the time reversal symmetry which is important because it
influences the bifurcations of periodic orbits.

%
The paper is organized as follows. In sec.~2 we take a global point of
view and discuss the topology of the surfaces of section induced by the
potential. The linearized Poincar\'e map is obtained in sec.~3, and the
results are used to calculate the stability of periodic orbits in sec.~4.
In the remaining sections we calculate the stability of short periodic
orbits for billiards with linear (sec.~5), 
quadratic or magnetic potential (sec.~6), 
and a rotating system, which models the 
restricted three body problem (sec.~7). 

\Goodbreak

\section{Billiards with potential}

Let the Hamiltonian of the free motion be given by
\[
H(\r,\P) = \frac{1}{2} (\P - \Ap(\r))^2 + U(\r).
\]
The velocity is given by $\v = \P-\Ap$. 
In principle the calculation could be done with an additional positive
definite ``metric'' in the kinetic energy,
but in the application we have in mind we only need the above form.
Let the configuration space of the free motion be $\Real^2$.
The energy surface of the billiard with potential is just the
ordinary energy surface but with $\r$ restricted to the interior of 
the billiard $\Omega$,
\[
        \Eh = \{ (\r,\P) \| H(\r,\P)=h, \r \in \Omega \}.
\]

Let the solution of Hamilton's equations
$\dot \XP = \J \nabla \XP$ be given by the flow $\phi$
\[
        \phi^t : \XP_1 \mapsto \XP(\XP_1;t),
\]
where we introduced $\XP=(\r,\P)$ and the symplectic matrix $\J$.
The section condition in our system is naturally given by the billiard
boundary $\partial \Omega \subset \Real^2$ or explicitly by 
$\r(t) = \R(s)$, such that the surface of section $\Sh$ becomes
\[ \label{eqn:sh}
        \Sh = \{ (\r,\P) \| H(\r,\P)=h, \r \in \partial \Omega \}.
\]
On the surface of section $\Sh$ we introduce coordinates
$\xi = (s,\p)$, where $s$ is the arc length of $\partial \Omega$
and $\p=\SP{\T}{\v}$ is the component of the velocity $\v = \P - \Ap(\r(s))$ 
parallel to the tangent vector $\T(s)$.\footnote{%
Note that the $\p$ stands for parallel component of the 
{\em velocity}, not of the momentum $\P$}
As we will see in the next
section this is a canonical coordinate system on $\Sh$ and
therefore $P$ is area preserving in these coordinates.
It is convenient to extend this coordinate system on $\Sh$ to 
the neighborhood of $\Sh$ in phase space. For this we take 
as local coordinates 
the energy $h$ in the direction of $\nabla H$, and the time
$\cau$ in the direction of the flow $\J \nabla H$.
The coordinate tuple $(s,\cau,\p,h)$ will be denoted by $\XX$.

The Poincar\'e map $P$ from bounce to bounce in a billiard 
with potential can now be described by a succession of four maps: 
\[
\begin{array}{rccll}
\sigma:         & \XX_1         & \mapsto & \XP_1 & 
  \mbox{the map from the surface of section to phase space, $\XX_1=(s_1,0,\p_1,h)$} \\
\phi^{\tau_1}:  & \XP_1         & \mapsto & \tilde \XP_2 & 
  \mbox{the flow map of free motion in the potential} \\
\isig:          & \tilde \XP_2  & \mapsto & \tilde \XX_2 &
  \mbox{the map back to local coordinates on the surface of section} \\
\refmap:        & \tilde\XX_2   & \mapsto & \XX_2 &
  \mbox{the reflection, $\XX_2 = (s_2,0,\p_2,h)$}
\end{array}
\]
The time $\tau_1$ must be chosen in such a way that $\tilde\XP_2 \in \Sh$,
where the tilde denotes the moment before reflection, see below.
As we will see, the description as a four dimensional mapping is
convenient for the linearization of the map. If we write $\sigma \xi$
we really mean $\sigma \Xi$ where $\Xi$ is the trivial 4-D extension of $\xi$,
similarly for $\isig$.

In this notation the Poincar\'e map reads
\[ \label{eqn:defPmap}
\begin{array}{rcl}
        P: \Sh & \rightarrow & \Sh  \\
                \xi & \mapsto & P(\xi) = \refmap \isig \phi^\tau \sigma \xi; \\
            & & \mbox{with $\tau$ determined by\ } \phi^\tau \sigma \xi \in \Sh.
\end{array}
\]
Note that if there is no reflection (i.e.\ if we study a Poincar\'e 
section of the free Hamiltonian) $\isig$ from the last iteration and
$\sigma$ from the next cancel, such that
the iterated map is given by 
\[
        P^n\xi_1 = \isig \phi^{\tau_n}\cdots\phi^{\tau_1}\sigma \xi_1 =
                \isig \phi^{(\tau_1 + \cdots + \tau_n)} \sigma \xi_1
\]
as must be the case.

The tilde on $\tilde\XP_2$ and $\tilde\XX_2$ indicates that we are on the surface of 
section but still in the instant before the reflection, 
such that the normal component 
$\tilde\n_2=\SP{\N_2}{\tilde\v_2}$ of $\tilde\v_2$ is smaller than 0.
Here we take the point of view described in \cite{DW95} that the surface
of section consists of both parts with either sign of $n_2$. 
In the case of an ordinary billiard with a smooth boundary
$\Sh$ becomes a 2-torus.
This 2-torus is divided into two cylinders by the lines where the flow 
is tangent to the surface of section. 
They are given by the condition
that the velocity $\v$ is parallel (or anti parallel) to the tangent $\T$,
or in local $\T$-$\N$-coordinates by $\n=0$.
In the case of a convex billiard these lines are two periodic orbits
given by the clockwise and counterclockwise sliding motion 
along the boundary. One cylinder carries initial conditions with the
velocity pointing inward (i.e.\ after the reflection, $\n>0$) while on the 
other one the velocity is pointing outward 
(i.e.\ before the reflection, $\tilde\n<0$).
The flow $\phi^\tau$ maps the first cylinder to the second,
and the reflection $\refmap$ maps the second back to the first.
In local $\T$-$\N$-coordinates $\refmap$ is trivial: it just changes 
$\tilde\n$ to $-\n$. This is the reason why we first use $\isig$ and then 
$\refmap$. Actually the points on $\Sh$ with outward pointing
velocities can be considered the same points as the ones with
the reflected inward pointing velocity, such that the reflection map
gives an identification of the boundary of the filled torus 
$\Omega \times S^1$, which is the energy surface, 
and turns $\Eh$ into a manifold without boundary. In the present case
we obtain $S^3$ by this construction.

If we add a potential the topological situation changes: the accessible 
region of motion in configuration space $\Qh$ can now be a subset of the 
region $\Omega$ bounded by the billiard walls. As usual $\Qh$ is given
by the condition $U(\r) \le h$, such that for the billiard with potential
we obtain
\[
        \Qh = \{ \r \| U(\r) \le h, \r \in \Omega \}.
\]
The additional boundaries $\partial \Qh \setminus \partial \Omega$ 
are the ovals of zero velocity \cite{Birkhoff27}, 
where the kinetic energy vanishes due to the additional 
potential. Moreover there might be a subset of the energy surface,
such that orbits in this subset never meet the boundary of $\Omega$. 
In the examples we are only dealing with integrable free Hamiltonians 
such that this part of the energy surface will be foliated by tori.
Although the Poincar\'e section will be $\Eh$-incomplete
\cite{DW95}, i.e.\ not every orbit in the energy surface will intersect
$\Sh$, the missing orbits are well under control and we ignore 
them in the following discussion. 
A very important point in adding a potential is that $\Eh$, 
the dynamics, and also $\Sh$ in general depend on the energy:
instead of the entire $\partial \Omega$ the part of the 
boundary in the accessible region of configuration space
can now become a collection of intervals $\I \subset \partial \Omega$ if
the energy is not too high.
On every point of $\I$ we have a certain amount of kinetic energy
available such that the norm of the velocity is fixed and
all possible directions can be parametrized by an angle,
such that we have to attach circles $S^1$ to every point of $\I$.
However, at the boundaries of $\I$ where the ovals of zero 
velocity intersect the billiard boundary $\partial \Omega$,
this circle has to be contracted to a point since the only possible
velocity is 0. Therefore the surface of section becomes a
collection of spheres, one for each interval in $\I$. On each sphere
there is a line where the velocity is tangent to $\partial \Omega$,
and it divides the sphere into two disks. If we restrict
our attention to the moment after reflection, the Poincar\'e map
will be defined on this collection of disks.

\section{The linearized map}

\def\fignotation{
The orbit $\r(t)$ encounters the boundary $\partial \Omega$ given
by the arc length parametrization $\R(s)$ at $\R$ with velocity
$\tilde \v$ before the reflection. The angle between the tangent $\T$ at
$\R$ and a fixed horizontal reference direction is $\alpha$. 
The angle between the reflected velocity $\v$ 
and the normal vector $\N$ is $\beta$. In the $\T$-$\N$-coordinate
system the velocity $\v$ is given by $(\p,\n)$, such that
$\p = |\v| \cos \beta$. Usually all quantities have an index to 
denote the point at which they are to be evaluated.
}
\def\FIGnotation{%
\centerline{\psfig{figure=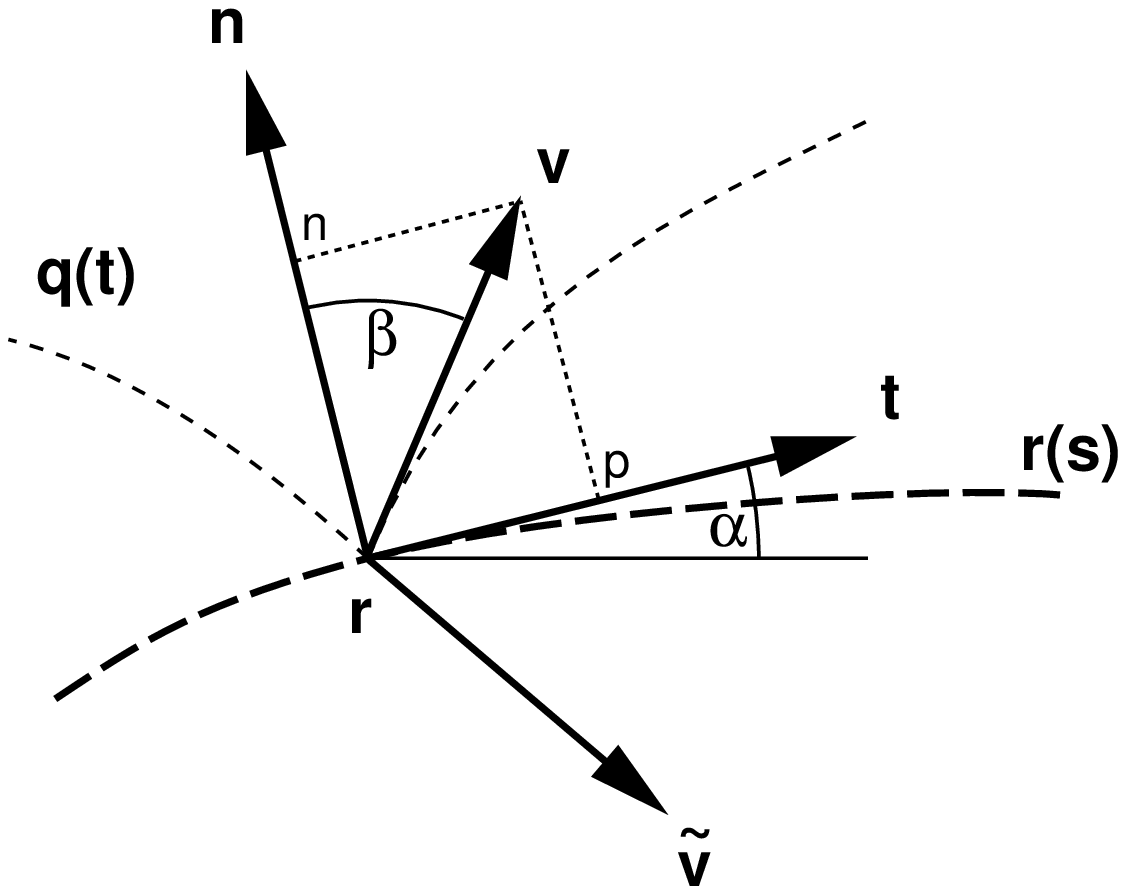,width=10cm}} }
\FIG{fig:notation}{\fignotation}{\FIGnotation}

For a general discussion of a Poincar\'e map from a (curved)
surface in the energy surface to itself see e.g.~\cite{AM78,Gutz90,MH92}.
As we will see, a symplectic coordinate system in the neighborhood of $\Sh$ 
is given by the arc length $s$, the parallel component $\p$ of the velocity 
$\v$ ({\em not} of the momentum $\P$), the value of the Hamiltonian $h$, and
the flow time $\cau$ starting at the point specified by $s,\p,h$.
Denote the components of this coordinate system by $\XX = (s,\cau,\p,h)$.
The transformation to the original canonical variables 
$\XP = (\r,\P)$ is given by
\[ \label{eqn:canotraf}
        \XP = \sigma\XX = \phi^\cau \F(s,\p,h),
\]
where $\F$ gives the initial point on $\Sh$ in the coordinates $\XP$.
The position is given by $\r = \R(s)$ and the velocity is
$\v = \D(\alpha(s)) \u$ where $\u = (\p,\n)$ is the velocity in the
local $\T$-$\N$-coordinate system and $\D=(\T,\N)$ the rotation
matrix.
$\alpha$ is the angle between $\T$ and the horizontal axis.
To fix an orientation of the $\T$-$\N$-system we let $\N$ always point
into $\Omega$ and then rotate by $\S=\D(\pi/2)$ to obtain $-\T=\S\N$.
By this convention the curve $\partial \Omega$ must be traversed 
counterclockwise for the billiard inside the curve and clockwise for
the billiard outside the curve. This automatically fixes the
sign of the curvature $\kappa$ to be positive for the billiard inside
a convex domain $\Omega$, and negative for the corresponding outer billiard.

The normal component $\n$ is implicitly defined via the shifted energy
function on $\Sh$ given by
\[ \label{eqn:Epn}
        \Epn(s,p,n,h) = \frac{1}{2} (\p^2 + \n^2) + U(\R(s)) - h,
\]
where the positive square root gives $\n$ and the negative $\tilde \n$.
Now $\F$ is given by
\[ \label{eqn:pis}
        \F : (s,\p,h) \mapsto \XP = (\r,\P)^t = (\R(s), \D(\alpha(s))\u + \Ap)^t.
\]
We only want to know the transformation (\ref{eqn:canotraf})
close to $\Sh$, i.e.\ for small $\cau$ such that we can approximate
$\phi^\cau \approx \Id + \cau \J \nabla H$, where 
the Hamiltonian vector field $\dot \XP = \J \nabla H$ at $\R(s)$ in the
coordinates $\XP$ is needed,
\[ \label{eqn:dotpis}
 \dot \XP = \vecII{\dot \r}{\dot \P} = \J \nabla H = 
	\vecII{\v}{\SP{\pr \Ap}{\v}- \pr U}.
\]
The complete transformation, which gives $\sigma$ for $\cau=0$, is given by
\[ \label{eqn:Xtrafo}
        \XP = \F(s,\p,h) + \cau \dot \XP|_{\F(s,\p,h)}
                = \vecII{\R + \cau \dot\r}
                        {\P + \cau \dot\P}
,
\]
with $\P$ and $\dot\P$ given by (\ref{eqn:pis},\ref{eqn:dotpis}) 
and $\n$ implicitly determined by (\ref{eqn:Epn}).
Now we want to find the linearization of this map at $\cau=0$ and show that 
it is a symplectic mapping. By this we find that the variables 
$\xi = (s,\p)$ are canonical on $\Sh$. Moreover we need exactly this
symplectic mapping to determine the linearized Poincar\'e map.
Using (\ref{eqn:Epn}) we obtain the derivatives of $\n$.
With the abbreviation $\nut = \SP{\pr U}{\T}$ we find
\[
\pafra{\n}{(s,\p,h)} = -\pafra{E}{(s,\p,h)} \Big/ 
                                \pafra{E}{\n} = 
                (-\gamma,-\p,1)/n.
\]
The Jacobian of the transformation (\ref{eqn:Xtrafo}) 
at $\cau = 0$ is the $4\times 4$-matrix
\begin{eqnarray}
\L = \left.\pafra{\XP}{\XX}\right|_{\cau=0}
        & = &   (\ps F          , \dot \XP      , \pp F         , \ph F ) \\
  & = & \matIIIV{\T             &  \v           & \Nu           & \Nu     }
                {\ps \P         &  \dot\P       & -\S\v/n       & \N/n} 
   =:  \matII{\A_l & \Nu }{\C_l & \D_l},
\end{eqnarray}
$\S = \D(\pi/2)$, and with (\ref{eqn:pis}) we find
\begin{eqnarray}
  \ps \P 
        & = & (\ps \D) \u + \D \ps u + \SP{\pr \Ap}{\ps \R} \\
\label{eqn:dsp}
        & = & \kappa \S \v - \N \nut / \n + \SP{ \pr \Ap}{ \T }.
\end{eqnarray}
The symplectic condition, $\L^t\J\L=\J$, 
states that $\A_l^t\D_l=\Id$ and that $\A_l^t\C_l$ be symmetric. 
The first holds
because of $\T^t\N=0$, $-\v^t\S\v=0$ and $\v^t\N=-\T^t\S\v=\n$, such that
$\D_l=\A_l^{-1t}$ and $\det \A_l = \n$.
The second condition is the identity
\begin{eqnarray}
               \SP{\v}{\ps \P}  & = & \SP{\T}{ \dot \P} \\
 -\SP{\v}{ \N} \nut / n + \SP{\v}{ \pr \Ap \T} & = & \SP{\T}{ \pr\Ap^t \v} - \SP{\T}{ \pr U} \\
        -\nut + \SP{\v}{\pr A \T} & = & \SP{\T}{\pr\Ap^t \v} - \nut.
\end{eqnarray}
The inverse of $\L$ can be simply computed because $\L$ is symplectic:
\[ 
        \label{eqn:LLinv}
        \L   =  \matII{\A_l & 0}{\C_l & \A_l^{-1t}} \qquad
        \L^{-1}  =  \matII{\A_l^{-1} & 0}{-\C_l^t & \A_l^t}.
\] 
To linearize the Poincar\'e map we use $\L|_{\xi_1}=\L_1$ to map vectors
at $\xi_1$ from $T\Sh$ to Euclidean coordinates, which are then mapped by
the \mojac\ of the flow $\M=\partial \phi^t/\partial \XP$.
At the new point $\XP_2$ we return to the section coordinates 
with $\L^{-1}$
and keep in mind that we must still perform a reflection.
In the local coordinate system this just means to put $-\n_2$ instead
of $\tilde\n_2$ in $\tilde\L_2^{-1}$. To indicate this we write
$\tilde \L_i = \tilde \L(s_i,\p_i,\n_i) = \L(s_i,\p_i,\tilde\n_i)
=\L(s_i,\p_i,-\n_i)$
such that
\[    \label{eqn:DQ}
        \DQ{2}{1} = \tilde \L_2^{-1} \M_1 \L_1.
\]
To obtain a $2\times 2$ matrix for $DP$ we just set $d\tau$ and $dh$ 
equal to zero, i.e.\ we consider the restriction of the map to the tangent 
space $T\Sh$ in local coordinates on $\Sh$. Deleting the appropriate
rows and columns from $\L$ and $\L^{-1}$ we find
\[ \label{eqn:DP}
        DP_{21} = 
        \matII{-(\S\tilde\v_2)^t/\tilde\n_2 & \Nu}{-\ps\P_2^t & \T_2^t}
          \M_1  \matII{\T_1 & \Nu}{\ps\P_1 & -\S\v_1/\n_1},
\]
where the left matrix is a $2\times 4$-matrix and the right is a
$4\times 2$-matrix.
For the trivial Hamiltonian of the ordinary billiard 
the only term in $\ps\P$ is related to the curvature, see
(\ref{eqn:dsp}).
In the presence of a potential $U$ there is an additional contribution in 
the form of $\nut = \SP{\pr U}{\T}$, and a similar term $\SP{\pr \Ap}{\T}$ 
for the vector potential.

The \mojac\ for the free motion in ordinary billiards is
\[ \label{eqn:Mfree}
        \M = \matII{\Id & \tau \Id}{\Nu & \Id},
\]
and the main simplifications in this case come from the fact that 
$\tilde\v_2 = \v_1$, and with (\ref{eqn:dsp}) we can rewrite
(\ref{eqn:DP}) as
\begin{eqnarray}
        DP_{21} & = & \matII{1&0}{-\kappa_2\n_2&1}
                        \matII{-\v_2^t\S/\n_2 & \Nu}{\Nu & \T_2^t}
                        \matII{\Id & \tau \Id}{\Nu & \Id}
                        \matII{\T_1&\Nu}{\Nu & -\S\v_1/\n_1}
                        \matII{1&0}{-\kappa_1\n_1&1}          \\
\label{eqn:DPoldlin}
& = &   \matII{\frac{1}{\n_2}&0}{0&\n_2}
        \matII{1&0}{-\frac{\kappa_2}{\n_2}&1}
        \matII{-1&-\tau}{0&-1}
        \matII{1&0}{-\frac{\kappa_1}{\n_1}&1}
        \matII{\n_1&0}{0&\frac{1}{\n_1}}
\end{eqnarray}
which is a well known formula, see e.g.~\cite{HenWis83,DRW96}.
For cases with potential more explicit formulas than (\ref{eqn:DP}) tend to be
quite complicated. 
As we will see, it is nevertheless possible in some cases to perform
the stability analysis up to period two.
The obvious use of (\ref{eqn:DP}) is in the numerical calculation of
Lyapunov exponents (see e.g.~\cite{DRW96,BD95}) and in the numerical search 
for periodic orbits with Newton's method. 
Note that if different, possibly non-canonical,
coordinates are used, as the arc length parametrization 
of $\partial \Omega$ may be too complicated,
the expression for $DP_{21}$ has to be sandwiched by the Jacobian
of this transformation for Newton's method to work.
Another application is the analytical estimation
of the Lyapunov exponent as performed in \cite{BD95}.
Let us remark that the linearized map
in the full phase space (\ref{eqn:DQ}), besides being conceptually
much clearer than (\ref{eqn:DP}), can be useful to implement
a continuation method to find orbit cylinders in phase space, i.e.\ the
collection of periodic orbits for different $h$, because it contains
information about the change of the orbit if the energy changes.

\section{Linear stability of periodic orbits}

The standard application of the linearized map is to calculate the 
stability of periodic orbits. There is one simplification in calculating
the stability of periodic orbits as compared to the calculation of 
the general linearized Poincar\'e map, namely that we are only interested in 
the trace of a product of linearized maps, and that we can cyclically 
permute matrices under the trace. 
The $4\times 4$ monodromy matrix $\MOf_n$ of a period $n$ orbit is given by
\[
        \MOf_n = \DQ{1}{n} \DQ{n}{n-1} \cdots \DQ{2}{1} 
\]
and using (\ref{eqn:DQ}) we obtain
\[
        \Trf_n := \tr \MOf_n = \tr \left
                (       \tilde\L_1^{-1}\M_n\L_n \, 
                        \tilde\L_n^{-1} \M_{n-1}\L_{n-1} \cdots
                        \tilde\L_2^{-1} \M_1\L_1
                \right ).
\]
Now we define the linearized reflection matrix
\[ \label{eqn:Kdef}
        \K_i = \L_i^{-1} \tilde \L_i,
\]
and with a cyclic permutation we find
\[ \label{eqn:trMK}
        \Trf_n = \tr \prod_{i=1}^{n} \M_i \K_i.
\]
The stability is split up into contributions from the reflection $\K$
and from the free motion $\M$. 
Moreover, both contributions depend
only on a single point, while for $\DQ{2}{1}$ we needed information about
the initial and final point.
A similar formula with $2\times 2$ matrices
for the ordinary billiard is well known, see e.g.~\cite{CRL90,SieberSteiner90}. 
Multiplying $DP_{32}DP_{21}$ as given by (\ref{eqn:DPoldlin}) we find
\[
        ...\matII{1&\tau_2}{0&1} \matII{-1&0}{2\kappa_2/\n_2&-1}...
\]
In (\ref{eqn:trMK}) we have the natural generalization for an arbitrary 
billiard with potential, which is conveniently
formulated in four dimensions. 

The reflection matrix defined in (\ref{eqn:Kdef}) can be written as
\[
        \K = \matII{\W&\Nu}{\CK&\W}.
\]
It is symplectic, which amounts to $\W\W^t = \Id$ and $\W^t\CK$ symmetric.
$\W$ is a symmetric involutive matrix with $\det\W=-1$ and
$\W^2 = \Id$, as must be the case for a reflection:
\[ \label{eqn:Wmat}
        \W = \matII     {\cos2\alpha &  \sin2\alpha}
                        {\sin2\alpha & -\cos2\alpha}.
\]
The eigenvalue-eigenvector pairs of $\W$ are 
$(+1,\T)$ and $(-1,\N)$. The reflection
line is $\T$ and the direction of reflection is $\N$, recall that
$\alpha$ is the angle between $\T$ and the horizontal axis.
Note that the reflection $\W$ acts on both, momenta and positions. 
In this way the mapping ``corrects'' the 
time at which the reflection took place:
The symplectic
coordinate system $\XX$ ensures that we do not have to bother to directly
calculate the (linear approximation of the) change of time
between reflections.

The lower left block $\CK$ of $\K$ is much more complicated than the 
other blocks and contains all the contributions from $\kappa$, $U$ and $\Ap$.
From (\ref{eqn:Kdef}) and (\ref{eqn:LLinv}) we find after some manipulations
\[ \label{eqn:Cis}
        \CK = 2 \frac{\kappa}{\n} \S\D\u \, (\S\D\W_0\u)^t + 
                \frac{2}{\n} \SP{\pr U}{\N} \N \N^t  - 
                \frac{4\p}{n}\Bmag \N \N^t + 4 \SP{\N}{\pr \Ap \T} \S,
\]
where $\W_0$ is the above reflection matrix at $\alpha=0$ and
with $\r = (x,y)$ we introduce
$ 
        \Bmag = \partial_x \Ap_y - \partial_y \Ap_x
$, 
the magnetic field.
Except for the skew symmetric contribution from $\pr \Ap$ the matrix $\CK$ is 
obviously singular. For constant magnetic fields 
we have $\SP{\N}{\pr \Ap \T} = \Bmag/2$.
In the following we will make this assumption, which is valid for the
examples treated below.
Note that $\S\D\W_0\u = \S\W\D\u = \S\W\v$ 
such that we obtain
\[
        \CK =   \frac{2\kappa}{\n} \S\v \, (\S\W\v)^t + 
                \frac{2}{\n} \SP{\pr U}{\N} \N \N^t  - 
                2 \Bmag (\frac{2\p}{n} \N \N^t + \S).
\]
In the stability formulas we will often encounter
\[ \label{eqn:trCK}
        \tr \CK = \frac{2}{\n}\left(
                \kappa(\p^2-\n^2) + \SP{\pr U}{\N} - 2 \p \Bmag \right),
\quad 
    \CKe_{21} - \CKe_{12} = 4\kappa\p - 4 \Bmag.
\]

The monodromy matrix $\MOf_n$ has eigenvalue 1 twice, with 
left eigenvector $\nabla H$ and right eigenvector $\J\nabla H$
\cite{MH92,EcWi91}.
In the $\XX$ coordinate system it has the special form
\[ \label{eqn:DPEVs}
\MOf_n = 
\left( \begin{array}{cccc}
                a & 0 & b & x \\
                x & 1 & x & x \\
                c & 0 & d & x \\
                0 & 0 & 0 & 1
        \end{array} \right) 
        = \tilde \L_1^{-1} \M_n \K_n \cdots \M_1 \K_1 \L_1,
\]
where $x$ stands for arbitrary coefficients, and the entries $a$, $b$, $c$,
$d$ give the linearized $2\times 2$ Poincar\'e map $DP^n$ in local 
coordinates. From $\det \MOf_n = 1$ we see once more that 
$\det DP = ad-bc = 1$, such that $\xi$ are area preserving coordinates for $P$.
Right now we are interested in the fact that
the spectrum of $\MOf_n$ is $\{1,1,\lambda,1/\lambda\}$ such that
\[
        \Trf_n = 2 + \lambda + \frac{1}{\lambda}.
\]
By calculating $\Trf_n$ we therefore obtain the eigenvalues of 
$DP^n$:
\[
        \trt_n := \tr DP^n = \tr \MOf_n - 2 = \Trf_n - 2
= \lambda + \frac{1}{\lambda}.
\]

\subsection*{Period two orbits in the ordinary billiard}

Let us now calculate $\Trf_2$ for a period two orbit of the billiard without 
potential, 
where $\M$ is given by $(\ref{eqn:Mfree})$. It will turn out that the
same $\M$ holds for the case of the gravitational billiard.
\begin{eqnarray} \label{eqn:Tr2}
        \Trf_2 & = & \tr ( \M_2 \K_2 \M_1 \K_1 ) \\
                & = & 4\cos(2\alpha_1-2\alpha_2) +
                (\tau_1+\tau_2) \tr(\W_2\CK_1 + \CK_2\W_1) + 
                \tau_1\tau_2 \tr \CK_2\CK_1
\end{eqnarray}
For a period two orbit (without potential) we have $p_1=p_2=0$,
$\n_1=\n_2=1$, $\tau=\tau_1=\tau_2$ and $\alpha_1 = \alpha_2 + \pi$, 
hence we obtain
\[
        \Trf_2 = 4 - 4\tau(\kappa_1+\kappa_2) + 4\kappa_1\kappa_2 \tau^2,
\]
which is a well known result, see e.g.~\cite{KT91}.
Before passing to the new application to gravitational billiards,
we show the stability diagram of this period two orbit.
It is convenient to introduce Greene's residue $\GR$ \cite{Greene79}
and its ``complement'' $\bar\GR$ defined by
\[
         \bar\GR = 1 - \GR = \Trf/4, 
\]
such that in the above case
\[ \label{eqn:bilp2}
        1-\GR = \bar\GR = (1-\tau\kappa_1)(1-\tau\kappa_2).
\]

Bifurcations with $\lambda=+1$, $\GR=0$ are called ``direct parabolic'' and 
bifurcations with $\lambda=-1$, $\bar\GR=0$ ``invers parabolic''. 
The direct parabolic bifurcation with $\lambda=1$ generically means the 
creation of periodic orbits ``out of nothing'' in a saddle-center bifurcation.
Thus, the continuous dependence of orbits on the bifurcation 
parameter is not guaranteed at this point.
Index preservation \cite{GMVF81} requires the simultaneous creation
of one orbit with index $+1$, which must be elliptic close to $\GR=0$,
and one orbit with index $-1$ which is always direct hyperbolic.
In the presence of symmetry (e.g.\ time reversal symmetry) 
a symmetric orbit may depend continuously on the bifurcation parameter,
and in this case one observes a symmetry breaking pitch fork 
bifurcation instead \cite{Rimmer83}.

At an invers parabolic bifurcation with $\lambda = -1$, $\bar\GR=0$ the 
periodic orbit depends continuously on the bifurcation parameter, 
and this is the most important difference as compared to 
the direct parabolic case.
The orbit has index $+1$, and there is no way 
to create an orbit with the same period $n$ and index $-1$ because $\GR=1$. 
Therefore the orbit must survive the crossing of the line $\GR=1$,
and turns from elliptic via invers parabolic to inverse hyperbolic.
However, considering the map $P$ iterated $2n$ times, an inverse hyperbolic
orbit of period $n$ looks hyperbolic, such that we have index $-1$ for
$P^{2n}$, and therefore we typically expect the creation of two stable orbits 
of period $2n$ with index $+1$ (or the destruction of two unstable ones). 
This is called period doubling bifurcation.
Of course there can be more complicated bifurcations at $\lambda =\pm1$,
and also additional resonances in the elliptic region, 
see e.g.~\cite{GMVF81,Meyer70}.

\def\figbilpii{
Stability diagram for period two orbits in the billiard
without potential. Contour lines of the residue $\GR$ (\ref{eqn:bilp2}) 
are shown. The elliptic region is shaded. 
Inside this region the lines with $\GR=1/2,3/4$ are drawn. 
In the unstable regions the integer valued residues are shown.
The bold dashed line indicates the stable period two orbit
in the family of billiards with elliptic boundary. Period two orbits of the
Sinai billiard are in the lower left quadrant, those of the stadium 
billiard, or the unstable ones in the ellipse, are in the upper right 
region with $\GR<0$.
With an appropriate change of the axes labeling this diagram can also
be used for period two orbits of the billiard in a constant magnetic field
(\ref{eqn:MagStab}) and for period one and period two orbits of the billiard
with a harmonic potential (\ref{eqn:OszStab1},\ref{eqn:OszStab2a},\ref{eqn:OszStab2r}).
}
\def\FIGbilpii{%
\centerline{\psfig{figure=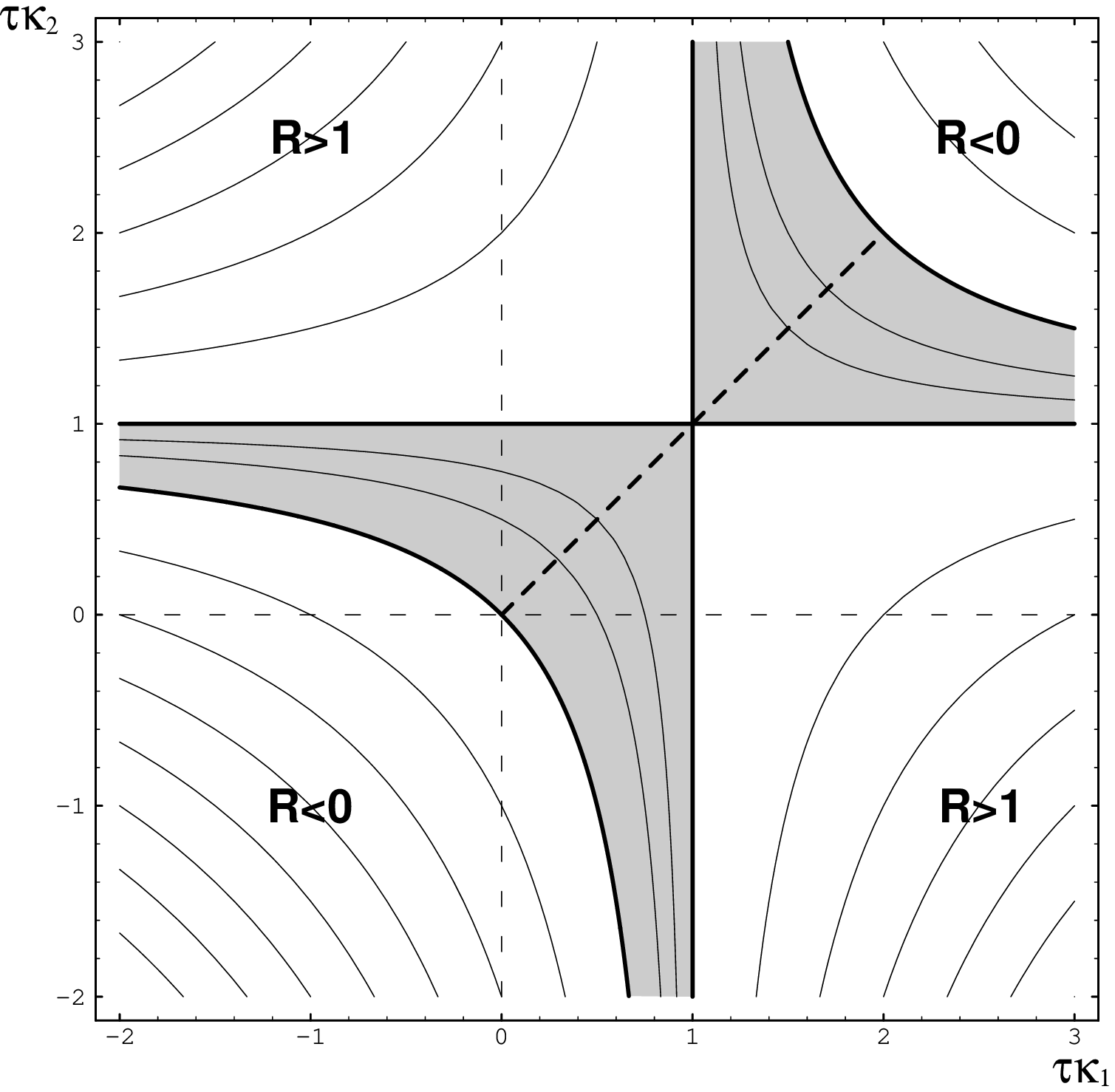,width=10cm}} }
\FIG{fig:bilp2}{\figbilpii}{\FIGbilpii}

In \fig~\ref{fig:bilp2} we show the level lines of $\GR$
as a function of $\tau\kappa_1$ and $\tau\kappa_2$. We could scale the length
by $\tau$ such that we always deal with a unit length orbit, but sometimes
it is more natural to think of an orbit as becoming longer instead of
the curvature becoming larger.
Regions with elliptic orbits ($0 < \GR < 1$) are shaded. The boundaries
are given by $\GR=0$ (i.e.\ by the hyperbola $\tau=1/\kappa_1+1/\kappa_2$),
the direct parabolic case,
and by the two lines $\bar\GR=0$ with $\tau\kappa_1=1$ or $\tau\kappa_2=1$,
the invers parabolic case.
The interpretation is as follows, where
due to symmetry we restrict the discussion to $\tau\kappa_2>\tau\kappa_1$:\\
If both boundaries at the reflection points are defocusing, $\kappa_i<0$, the 
orbit is unstable. The direct parabolic reflection at two parallel straight 
lines is at the upper right corner of this quadrant. 
If there is one focusing and one defocusing reflection the orbit is
usually inverse hyperbolic; however, even for large defocusing curvature,
elliptic motion is always possible.
If we lower the value of $\tau\kappa_2$ 
starting in the inverse hyperbolic region we will encounter an inverse 
parabolic bifurcation at $\tau\kappa_2=1$.
Then there is a small elliptic region which we leave via a direct  
parabolic bifurcation.
There is no further bifurcation if the boundary becomes defocusing.

The region with two focusing reflections, $\kappa_i>0$, is the 
most complicated.
If the orbit is sufficiently long, then it is direct hyperbolic, $\GR<0$.
This is the longer period two orbit that is always present in a smooth convex 
billiard. If we make the orbit shorter we encounter a direct parabolic 
bifurcation if $\tau\kappa_1$ is sufficiently close to 1.
Examples of orbits in the intermediate stable region are the
shorter period two orbit of smooth convex billiards (if they are stable).
At $\tau\kappa_1=1$ these orbits become inverse hyperbolic if the curvatures
are different, even though both reflections are focusing.
However if the length is decreased further the orbit becomes stable
again at $\tau\kappa_2=1$.

The period two orbit of the circle billiard is at 
$\tau\kappa_1 = \tau\kappa_2 = 2$,
and for an ellipse with semi major axes $a$ and $b$, $a > b$ the unstable
orbit along the longer axis is at $\tau\kappa_i=2a^2/b^2$ while
the stable one is at $\tau\kappa_i=2b^2/a^2$, indicated by the bold 
dashed line. For $a=b\sqrt{2}$ it is
invers parabolic, such that this situation is particularly susceptible to
perturbations. Note that the passage of the stable periodic orbit of
an integrable system through the special point where the two regions of
elliptic motion meet, $\kappa_1=\kappa_2$,
avoids additional bifurcations. We will see the
same behavior in the integrable billiard with parabolic
boundary in a gravitational field.

\Goodbreak

\section{The gravitational billiard}

In the gravitational billiard we have a linear potential $U = \SP{\g}{\r}$
with $\g = (0,g)^t$. Using the \mojac\ $\M$ of the free motion
\[
        \M = \pafra{\XP(t)}{\XP_1} = \matII{\Id & t\Id}{\Nu & \Id}
\]
the general solution is given by
\[
        \XP(t) = \M \vecII{\r_1}{\P_1} - \vecII{t^2\g/2}{t\g}.
\]
$\M$ is the same as for the ordinary billiard without potential. 
In this system a period one orbit exists if there is a point on the 
boundary with horizontal tangent, such that $\r(\tau)=\r_1$ and 
$\P(\tau)=-\P_1$. Their trace is given by
\[
        \Trf_1 = \tr \M \K = \tr (2 \W + \tau \CK) = \tau \, \tr \CK
\]
and with $\T=(1,0)$, $\N=(0,1)$ and $\p=0$ we obtain from (\ref{eqn:trCK}) 
that
\[
        \Trf_1 = 2\tau(-\kappa \n+g/n) = 4 - 2 \kappa \tau \n.
\]
In the latter equations we used $\tau=2\n/g$ to eliminate $g$ from the 
expression. With $h = \n^2/2$ and eliminating $\tau$ we find 
$\trt_1 = 2 - 8\kappa h /g $, which is well known, see e.g.\
\cite{KL91,KT91}. Additional relations 
like $\tau g = 2 n$ in this simple example, 
exist between the quantities entering the stability formulas, because we
have circumvented the use of the solution of the equations of motion.
The additional relations appear in the monodromy matrix 
in the $\XX$ coordinate system: It must have the form (\ref{eqn:DPEVs}),
i.e.\ the 0's and 1's are always there. 
In more complicated applications
it might be more appropriate to calculate these relations directly, but 
the final check with (\ref{eqn:DPEVs}) should always be made.

The interpretation of the bifurcation of the periodic orbit is as follows:
If the boundary is defocusing, as e.g.\ studied in \cite{Henon88},
the orbit is direct hyperbolic. 
At a flat boundary we have a direct parabolic bifurcation.
For a focusing boundary 
the orbit is elliptic for small energies and becomes inverse
hyperbolic if the energy becomes large compared to the radius of curvature.
The invers parabolic transition at $g=2h\kappa$ typically creates 
an elliptic period two orbit in a period doubling bifurcation.
Now we turn to the calculation of the stability of the period 
two orbit created in this bifurcation. We only treat the case of the
time and space symmetric period two orbit. The spatial symmetry is a
reflection at the vertical axis, such that the angles $\alpha_i$ between
the tangent at the reflection points and the horizontal axis 
are related by $\alpha_2 = -\alpha_1$. The time reversal symmetry
requires $\p_1=\p_2=0$.

Instead of calculating the full symmetric periodic two orbit
we introduce an upright wall 
at the place of the vertical symmetry axis
for which again $\p_2=0$ must hold. The reflection matrix $\K_2$ for
this wall with $\alpha_2 = \pm\pi/2$ and $\kappa_2 = 0$ is given by
\[
        \K_2 = \matII{\W_0 & \Nu}{\Nu & \W_0}.
\]
Formula (\ref{eqn:Tr2}) is also valid for the
gravitational billiard because it has the same \mojac\ as the
ordinary billiard, and it reduces to
\[
     \frac{\Trf_2}{4} = \cos(2\alpha\pm\pi) + \frac{\tau'}{2} \tr(\W_0\CK_1)
      = -\cos2\alpha +  \frac{\tau'}{n} \cos2\alpha(\kappa n^2 + g \cos\alpha).
\]
The time $\tau'$ is just half the time of the full orbit.
Using the relation $g\tau' = n\cos\alpha$
obtained from the explicit solution with the above boundary values,
we find 
\[ \label{eqn:grap2}
        1-\GR' = \bar\GR' = \cos2\alpha(\n\kappa \tau'  - \sin^2\alpha)
\]
The residue for the full orbit is now obtained from
\[ \label{eqn:factori}
        1 - \GR = \bar\GR = (1-2\bar\GR')^2.
\]
The reason that the residue (or its complement) factorize in the above 
fashion is usually related to a symmetry in the system. 
If we shift the zero of the potential energy to the initial point we
can again use $h=\n^2/2$ to eliminate $\tau$.

\def\figgrapii{
Stability diagram for the symmetry reduced period two orbit in 
the gravitational billiard. The contour lines of the residue 
$\GR' = 1-\bar\GR'$ (\ref{eqn:grap2}) are shown.
The elliptic region is shaded. 
Inside this region the lines with $\GR'=1/2,3/4$ are shown. 
In the unstable regions the integer valued residues are shown.
The dashed line indicates the stable periodic orbit
in the family of billiards in the paraboloid. The dotted line 
corresponds to the wedge billiard and the long dashed line to 
the circle billiard.}
\def\FIGgrapii{%
\centerline{\psfig{figure=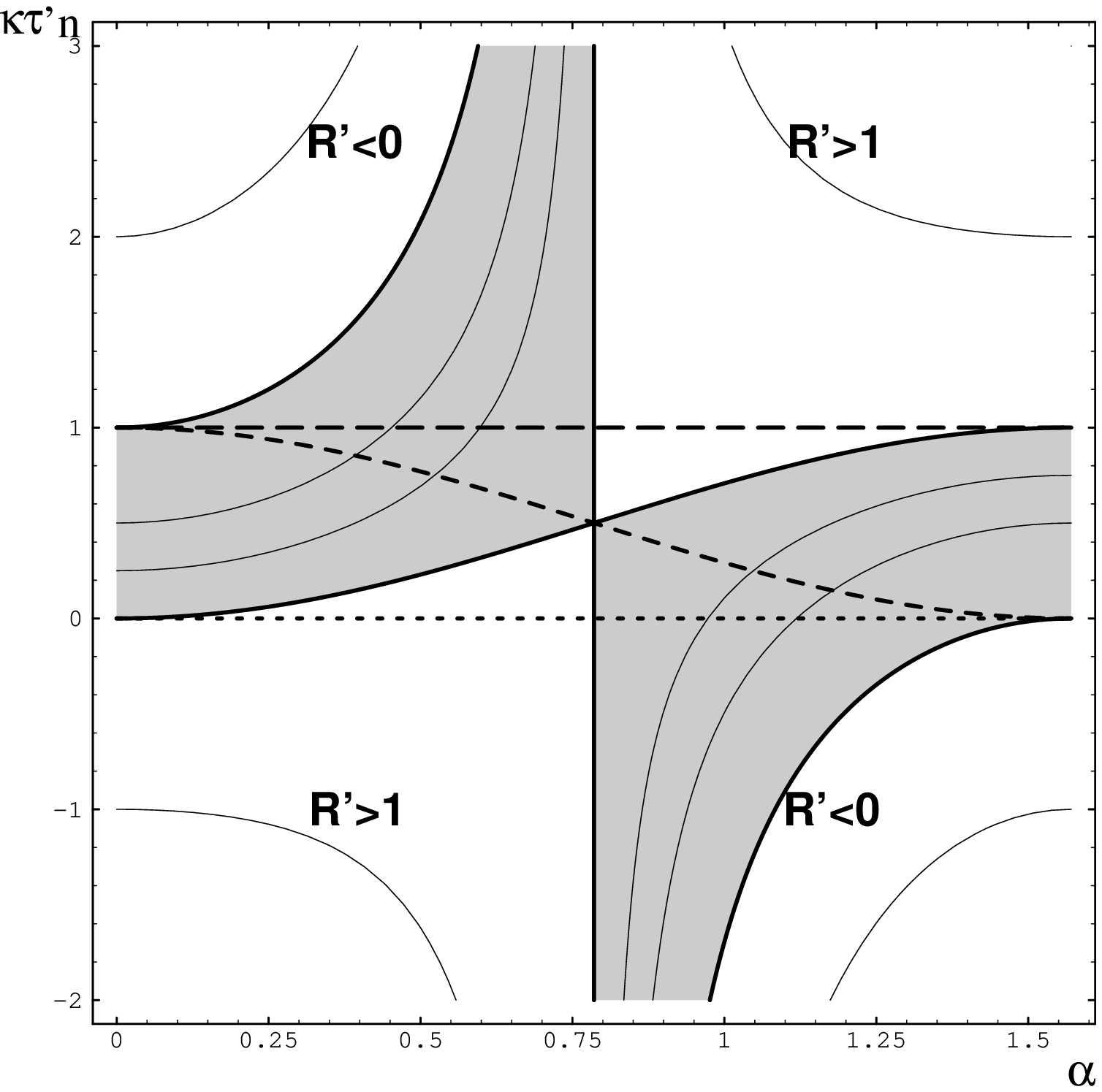,width=10cm}} }
\FIG{fig:grap2}{\figgrapii}{\FIGgrapii}

Let us now study the stability of this periodic orbit in a diagram 
similar to \fig~\ref{fig:bilp2}. Recall that we are not discussing
the question of existence of the orbit. In fact this orbit (even in
symmetric boundaries) does not exist for all energies. But if it exist,
its stability can be read off \fig~\ref{fig:grap2},
where the residue $\GR' = 1-\bar\GR'$ 
of the symmetry reduced orbit is plotted as a function of 
$\n\kappa\tau'$ and $\alpha$.
We start
by analyzing classes of orbits with constant angles $\alpha=0,\pi/2,\pi/4$.

For $\alpha=0$ we have the orbit bouncing up and down on a vertical
line. This is not, however, the period one orbit discussed above, because
there is also an upper wall. The residue is just $1-\n\kappa \tau'$.
For infinite energies this line describes periodic orbits
of the billiard without potential. To compare to \fig~\ref{fig:bilp2}
we must pass to the unreduced system with $\GR = 4\GR'(1-\GR')$
and $\tau = 2\tau'$. Then the line $\alpha=0$ can be mapped to the
diagonal of \fig~\ref{fig:bilp2}.

In the other extreme we have $\alpha=\pi/2$ which means a reflection
at a vertical wall with $\GR'=\n\kappa \tau'$. This motion can only be
realized for infinite energies or vanishing distance and we can connect 
this line to the period two orbit in the billiard without potential:
the residue of the full system on the line $\alpha=\pi/2$ can again
be mapped to the diagonal of \fig~\ref{fig:bilp2}.

There is the surprising feature that at $\alpha=\pi/4$ the orbits
are invers parabolic, {\em independently} of all the other quantities.
For an intuitive explanation consider the non-symmetry reduced system:
to throw a ball
the farthest distance with least energy it must take off with an angle
of $\pi/4$. For higher energies and the same distance there are two
distinct solutions, the higher orbit taking a longer time than the
lower one. Together these two orbits together give a period two space symmetric
orbit which does not posses time reversal symmetry. 
In the full system the time symmetric orbits have residue 0, 
and we have 
a symmetry breaking pitch fork bifurcation,
which appears as a period doubling bifurcation in the reduced system.

Let us now look at orbits with arbitrary angle $\alpha$ and defocusing
boundary, $\n\kappa\tau' < 0$.
For $\alpha < \pi/4$ the orbits are always inverse 
hyperbolic (in the reduced system). Note that in the full system
the symmetric period two orbit is always direct hyperbolic if it is
unstable. If the angle is increased beyond $\pi/4$ the defocusing
boundary orbit always becomes first elliptic and eventually direct
hyperbolic as the ordinary billiard at $\alpha = \pi/2$ is approached.
For a focusing boundary ($\n\kappa \tau' > 0$) the orbit is (almost) always 
stable at $\n\kappa \tau'=1/2$. We observe that the whole picture 
is invariant under a rotation by $\pi$ around the point $(\pi/4,1/2)$.
With this duality we can translate the statement 
``defocusing boundary for small angles is inverse hyperbolic'' to
``sufficiently focusing boundary and large angles is inverse hyperbolic'',
and similarly ``sufficiently focusing boundary and small angle is hyperbolic''.

Now we discuss the evolution of this orbit in specific examples.
The creation of the period two orbit by period doubling from the upright
bouncing motion happens at $\alpha=0$, $\n\kappa \tau=1$ with residue 0.
This point is part of two well known billiards, the integrable parabola 
billiard \cite{KT91,KL91} and the (gravitational) billiard inside 
the circle \cite{HayVid94}. The line of the period two orbits of these systems,
parametrized, e.g., by the energy, is also given in \fig~\ref{fig:grap2}.
In the circle billiard $\n\tau$ is constant ($\kappa$ is trivially a constant).
For small energies (and therefore angles) the orbit is stable, 
at $\alpha=\pi/4$, it becomes unstable in a period doubling
bifurcation and stays unstable henceforth. At infinite energy the
billiard becomes the integrable circle billiard without potential
and the period two orbit is direct parabolic.
For the parabola billiard the behavior is quite different. The creation
of the orbit starts at the same place, but according to the already 
stated observation that integrable systems avoid bifurcations it 
stays in the elliptic region when the angle and energy are increased,
passes to the symmetry point of the diagram and stops at the
infinite energy limit.
The third line shown in \fig~\ref{fig:grap2} corresponds to the
period two orbit in the wedge billiard \cite{LeMi86,RSW90}, for which
the curvature is zero. At $\alpha=\pi/2$
this line joins the one of the parabola billiard. Here, however, we do
not have an infinite energy limit but instead zero distance: The angle
$\alpha$ is also the wedge angle, which is the only system parameter, 
and at $\pi/2$ the wedge degenerates into a line. For $\alpha > \pi/4$
the orbit is elliptic, at $\pi/4$ where the billiard becomes integrable
it is invers parabolic and for smaller angles it is inverse hyperbolic, as
must be the case because the billiard is ergodic for $\alpha<\pi/4$
\cite{Wojtkowski90}.

Let us finally remark that the above stability formula for the full orbit
is quite similar if the curvatures are different at the reflection points,
$\bar\GR = (1-2\bar\GR'_1)(1-2\bar\GR'_2)$, where $\bar\GR'_i$ 
is given by (\ref{eqn:grap2}) with the
curvature $\kappa$ replaced by $\kappa_i$.

\section{Billiards with constant magnetic field or harmonic potential}

The stability formulas
for the symmetric period two orbits 
in the billiard with harmonic potential or magnetic field
are structurally similar to 
those of the ordinary billiard. In the symmetric case of the ordinary
billiard we find from
(\ref{eqn:bilp2}) that $\GR = 2 \kappa l (1-\kappa l/2)$ where
$l$ is the length of the orbit. The factor $1-\kappa l/2$ will
turn out to be present in the following examples with potential, such
that its geometric significance is independent of the additional
potential: If the Euclidean distance of reflection points is equal to the 
diameter of a circle with radius 
(of curvature) $1/\kappa$ the stability changes.

\subsection*{Constant magnetic field}

The vector potential of the constant magnetic field $\Bmag$ is
$\Ap = (-y\Bmag/2,x\Bmag/2)$, where
we denoted the components of $\r$ by $(\qrx,\qry)$.
The free motion takes place on circles with Larmor radius $|\v|/|\Bmag|$.
The \mojac\ of the free motion is given by
\[
        \M(t) = \matII{ (\Id+\D(\Bmag t))/2 & 
                        -(\Id-\D(\Bmag t))\S/\Bmag   }
                   {    \Bmag(\Id-\D(\Bmag t))\S/4 &
                        (\Id+\D(\Bmag t))/2       },
\]
where $\D(\Bmag t)$ is a rotation matrix, 
which will always be written with
its argument to distinguish it from the rotation $\D$ around the
angle $\alpha$.
The general solution is just given by $\XP(t) = \M(t) \XP(0)$.
We take $\Bmag < 0$ such that the particles rotate counterclockwise.

Since the circles of the free motion are already periodic orbits
the billiard does not have period one orbits.
For a period one orbit we need a free orbit that
returns to the initial point in configuration space, but with 
different velocities. The billiad wall is then adjusted in such a 
way that the final velocity is reflected to the initial one.
We again restrict 
ourselves to a symmetric period two orbit, i.e.\ $\alpha_2 = \alpha_1 + \pi$.
Because of isotropy of space we can set $\alpha_1 = 0$ and
$\alpha_2 = \pi$ (the same argument could have been used for the
billiard without potential), such that $\W_2 = \W_1 = \W_0$,
where the last index denotes the standard reflection matrix
($\alpha = 0$ in (\ref{eqn:Wmat})),
not the position of reflection.
Using the relation $2\beta=|\Bmag| t$, where $2\beta$ is the angle of the
part of the circle traveled by the free orbit to eliminate the time,
such that $\p = |\v|\cos\beta$ and $\n = |\v|\sin\beta$, we find
\[ \label{eqn:MagStab}
        \bar\GR = (1 + 2p/\Bmag\kappa_1)(1 + 2 p/\Bmag\kappa_2).
\]
Noting that the Euclidean distance $l$ between the reflection points is just
$l=|2\cos\beta|\v|/\Bmag| = |2\p/\Bmag|$, 
and that $\Bmag < 0$  we obtain the result given 
in \cite{RobnikBerry85} for $\beta < \pi/2$. 
The above formula is also valid for the case of the 
outer billiard (e.g.\ outside the sphere, see \cite{Berglund94}) with 
$\beta > \pi/2$. The stability diagram is the same as for the billiard 
without magnetic field given in \fig~\ref{fig:bilp2}.

\subsection*{Harmonic oscillator}

Let the potential be given by 
$U(\r)=\omega^2 \qrx^2/2 + \nu^2 \qry^2/2$.
The \mojac\ of the flow is
\[
        \M(t) = \matII{ \cC & \Oms^{-1}\cS }{ -\Oms\cS & \cC},
\]
with 
$\cC = \diag( \cos\omega t, \cos\nu t)$,
$\cS = \diag( \sin\omega t, \sin\nu t)$, and 
$\Oms = \diag(\omega, \nu)$. The general solution is 
$\XP(t) = \M(t) \XP(0)$. 
For a repulsive potential $\omega,\nu$ are imaginary which leads to 
$\cosh$ and $\sinh$ instead of $\cos$ and $\sin$.
This system can have many different period one orbits. 
For periodicity we require $\r(\tau) = \r_1 = (\qrx_1,\qry_1)^t
\not = 0$ 
where $\tau$ is the (as yet undetermined) period of the orbit.
The condition for a period one orbit reads
\[ \label{eqn:oszi1}
        \r_1 = \cC \r_1 + \cS\Oms^{-1} \P_1.
\]
Excluding the special orbits with $\det(\Id - \cC) = 0$ respectively
$\det\cS=0$, we can solve
for $\r_1$.
Inserting into the corresponding solution 
for $\P(\tau)$ gives the final momentum after time $\tau$ again at 
the position $\r_1$,
\[
        \P(\tau) = ( -\cS \Oms (\Id - \cC)^{-1} \cS \Oms^{-1} + \cC ) \P_1
                = -\P_1.
\]
Because of $\P(\tau) = -\P_1$ these orbits are self retracing: 
they go up the potential
until they reach the oval of zero velocity and then return the same path
in configuration space until they reach the initial point with reversed
velocity. In order to obtain a period one orbit of the billiard the
boundary at $\r_1$ must be perpendicular to $\P_1$.
This is the standard situation for billiards without magnetic 
field and we already encountered it for the period one orbit in the
linear potential. By construction these orbits have time reversal symmetry.
The normal $\N$ at the point of reflection is parallel to $\P_1$ and 
$\p = 0$. Taking the trace of $\K \M$ we find
\[
        \Trf_1 = 2\cos2\alpha(\cos\omega\tau - \cos\nu\tau) + 
                \frac{\CKe_{11}}{\omega} \sin\omega\tau + 
                \frac{\CKe_{22}}{\nu} \sin\nu\tau 
\]
with $\CKe_{ii}$ given by (\ref{eqn:Cis}). For the calculation of $\CK$ note
that $\N$ is eigenvector of $\W$ and $\v \parallel \N$, such that 
$\S\W\v = -\S\v$. 
From (\ref{eqn:oszi1}) we obtain
\[ \label{eqn:Op1}
        \P_1 = (\qrx_1\omega\tan\frac{\omega\tau}{2},
                \qry_1\nu\tan\frac{\nu\tau}{2})^t.
\]
The relation between energy and period for fixed initial and final
point $\r_1$ is a transcendental equation,
\[
        h = \frac{\qrx_1^2\omega^2}{2\cos^2\omega\tau/2} + 
        \frac{\qry_1^2\nu^2}{2\cos^2\nu\tau/2},
\]
and therefore we leave everything parametrized by the period $\tau$.
The divergence of the energy as a function of $\tau$ is related
to the fact that the period of oscillations in the harmonic oscillator
is independent of the amplitude. The only variation in $\tau$ comes
from the fact that we do not start at the origin. If the energy is 
large, the difference in $\tau$ due to this initial offset is relatively
small and for infinite energy $\tau$ approaches half the period 
$\pi/\omega$ or $\pi/\nu$.

Now we study the orbits whose reflection condition 
$\SP{\T_1}{\P_1} = 0$ with $\P_1$ given by (\ref{eqn:Op1})
is fulfilled
independently of $\tau$, i.e.\ for which $\qrx_1=0$ or $\qry_1=0$.
With an attractive potential the periodic orbits with small $\tau$
run upwards in the potential. For $\omega\tau$ or $\nu\tau > \pi$ the initial
momentum in the corresponding direction changes its sign, and the orbit goes
down first, then up on the other side and returns to $\r_1$. Periods
$\tau$ larger than the full period don't give periodic
orbits in the billiard.
In a repulsive potential the orbit goes up in the direction of the
origin because otherwise it would never return.
The outer billiard with repulsive potential is a scattering 
system without periodic orbits.
In the case $\qry_1=0$ and $\qrx_1>0$ we obtain
\[
\Trf_1 = 4\left(
        1 \pm \frac{\omega}{\nu}\kappa \qrx_1
        \tan\frac{\omega\tau}{2}\tan\frac{\nu\tau}{2}
   \right) \cos^2\frac{\nu\tau}{2},
\]
where $\pm$ is the sign of the initial momentum.
A similar expression holds for the case $\qrx_1=0$.

In the case of the isotropic harmonic oscillator with 
$\omega = \nu$ we get the simple expression
\[
        \Trf_1 = 4\cos^2\omega\frac{\tau}{2} \pm 
                4\qrx_1\kappa \sin^2\frac{\omega\tau}{2}.
\]
Introducing the energy $h$ and $U_1 = U(\r_1)$ gives
\[ \label{eqn:OszStab1}
        \GR = (1-\frac{U_1}{h})(1 \mp \kappa\qrx_1).
\]
In the case that $h=U_1$, this describes a degenerate orbit 
which does not move at all.
Let us first consider orbits with positive initial momentum.
For larger $h$ the orbit moves an increasing distance radially outward.
If the boundary is sufficiently defocusing ($\kappa \qrx<-1$)
the orbit is always hyperbolic, while for 
slightly defocusing boundary with $-1 < \kappa \qrx < 0$ 
it is always elliptic.
For a focusing boundary the orbit is stable for small energies and
inverse hyperbolic for large energies.
Note that the stability can be read off \fig~\ref{fig:bilp2}
with an appropriate reinterpretation of the axes.
For orbits with initially negative $x$-momentum the sign of $\kappa$ is 
defined with respect to the opposite normal, such that both
orbits with either sign of the initial momentum have the same
stability, despite the fact that one ``sees'' a focusing and
the other one a defocusing boundary.

In a repulsive potential we have $U_1<h<0$.
The upper bound on $h$ ensures that the oval of zero velocity 
around the origin (which is needed for the period one orbit) does not vanish.
We obtain the same stability formula, however now the minus sign
in front of $\kappa$ describes the (only possible) orbits with
initial negative momentum.

The symmetric period two orbits can also be easily calculated,
compare to the results in \cite{KT91}.
We start at $(\qrx_1,0)$, $\qrx_1>0$ always with negative initial 
momentum until we reach $(-\qrx_1,0)$. Due to the assumed symmetry we
can introduce a reflecting wall with curvature zero in the middle
and for the symmetry reduced orbit in the attractive potential we find
\[ \label{eqn:OszStab2a}
        \GR' =  (1-\frac{U_1}{h}) (1-\kappa l / 2)
\]
and $\GR = 4\GR'(1-\GR')$ for the full orbit with $l=2\qrx_1$ the
distance between the reflection points. 
The factor $1-U_1/h$ is
always positive and smaller than 1.
Again the stability can be read off \fig~\ref{fig:bilp2}.
For the repulsive potential we get the same formula, however, now
it gives the complement of the residue for the symmetry reduced orbit
\[ \label{eqn:OszStab2r}
        \bar\GR' =  (1-\frac{U_1}{h}) (1-\kappa l / 2)
\]
and we obtain $\GR = 4\bar\GR'(1-\bar\GR')$ for the full orbit.
The existence condition in the repulsive case is $h>0$, such that
the orbit can cross the potential hill at the origin.
Therefore $1-U_1/h$ is always larger than 1.

P.~Stifter \cite{Stifter96} has shown that the ordinary billiard
in the cardioid, which is ergodic \cite{Wojtkowski86}, is equivalent to the
billiard with repulsive potential $-\qrx^2 - \qry^2$ inside the
unit circle with center $(1,0)$ for the energy $H=0$. This is 
exactly the energy where the oval of zero velocity vanishes and
there is an unstable equilibrium point at $(0,0)$. Changing the
energy gives a system which does not correspond to the cardioid billiard.
Instead this gives an interesting example where a bounded system,
which is close to integrable for low energies $H\approx -4$ and 
high energies, is proven to be ergodic for the special intermediate
energy value $H=0$.

\section{The rotating billiard}

A rotating billiard, as introduced in \cite{MBLTJS95}, consists of 
an ordinary inner or outer billiard boundary that moves on a circle
around the origin. If we study the outer billiard with two discs the
system shows a striking similarity to the restricted three body problem, 
because in the latter system orbits with close encounters to either large 
body are effectively reflected, even though the gravitational force is
attractive, such that the hard core potential of the billiard can model 
this motion. This special case of
a rotating billiard will be referred to as the restricted three
body billiard, abbreviated as R3BB.
The Hamiltonian in a uniformly rotating frame of reference is given by
\[ \label{eqn:RotHam}
        H = \frac{1}{2}\P^2 - \omega(\qrx p_y - \qry p_x) 
                = \frac{1}{2}(p_x + \omega \qry)^2
                 + \frac{1}{2}(p_y - \omega \qrx)^2 - \frac{1}{2}\omega \r^2
\]
such that we have a harmonic repulsive potential $U = -\omega\r^2/2$
and a vector potential $\Ap = (-\omega \qry, \omega \qrx)^t$.
The \mojac\ of the free motion is
\[ \label{eqn:RotJac}
        \M(t) = \matII{ \D(\omega t) & t \D(\omega t)}
                        { \Nu & \D(\omega t) }, 
\]
such that the general solution can be written as $\XP(t) = \M(t) \XP(0)$. 
In the context of the restricted three body problem the analog of the
above Hamiltonian is referred to as the Jacobi integral; we will stay
with the former name and also call the value of the Hamiltonian the energy,
which is {\em not} the energy of the particle in the non rotating frame.
In rotating coordinates the discs
are fixed on the $\qrx$-axis at some distance from the origin. 
The straight line
motion looks rather (nice and) complicated in the rotating frame of reference. 
The free motion is circle-like due to the coriolis field but it is somewhat 
distorted due to the repulsive harmonic potential centered at the origin.
Depending on the value of the Hamiltonian there can be up to an
infinite number of period one orbits coexisting. 
Now we are going to calculate the stability of these orbits.

\def\figrotpi{ 
Stability diagram for the period one orbits in 
the R3BB as a function of $\kappa\qrx$ and $\omega\tau$ as given by the
residue $\GR$ in (\ref{eqn:rotp1}).
The elliptic region is shaded. 
Inside this region the lines with $\GR=1/2,3/4$ are shown. 
In the unstable regions the residues $-1$, $\pm2$, $\pm3$, $\pm5$, $\pm10$,
$\pm25$ and $\pm50$ are shown.
The typical geometric situation is to the left of the vertical line 
$\kappa\qrx=1$, where the boundary is defocusing or 
only slightly focusing. We consider this region now:
The horizontal lines indicate saddle center bifurcations. 
The retrograde orbit created 
at $\omega\tau \approx 4.08$ is stable for slightly larger period,
similarly the other retrograde orbits  with $\omega\tau \approx (2m+1+1/2)\pi$,
for which $h>0$.
The direct orbits ($h<0$) bifurcating at $\omega\tau \approx (2m+1/2)\pi$ 
are stable for slightly smaller period.
The infinite residues at $\omega\tau = 2 m \pi$ prevent the smooth 
continuation of a retrograde orbit to another retrograde orbit, similarly
for the direct orbits.
In the left two pictures the energy $h$ and the
angle $\theta$ between in- and outgoing velocities at the reflection point
as a function of $\omega\tau$ are shown.
}
\def\FIGrotpi{%
\centerline{
\hbox{
\psfig{figure=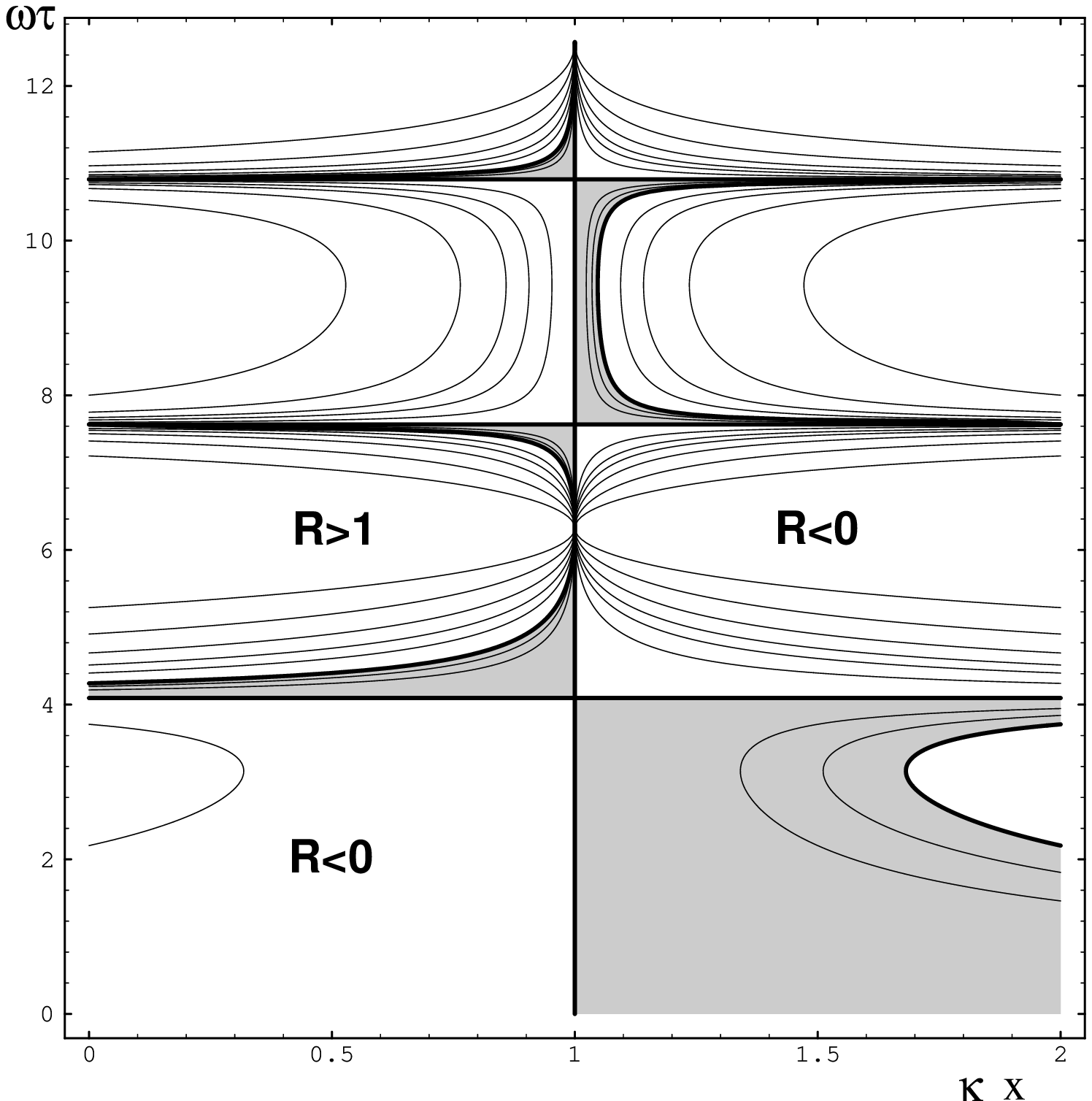,height=8cm}
\psfig{figure=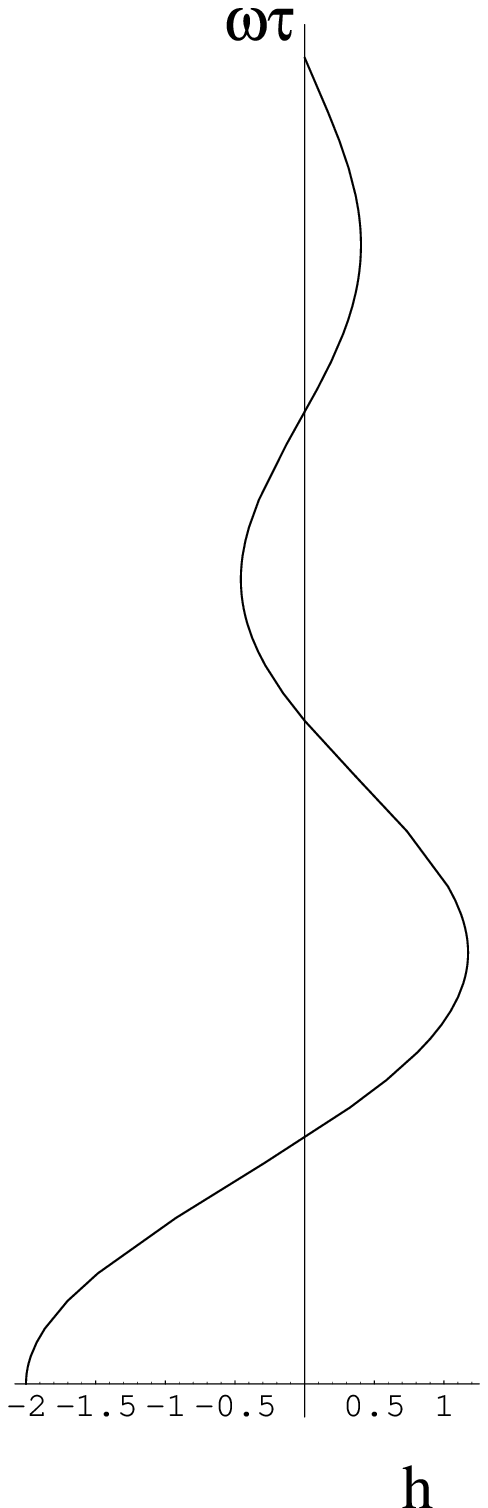,height=8cm}
\psfig{figure=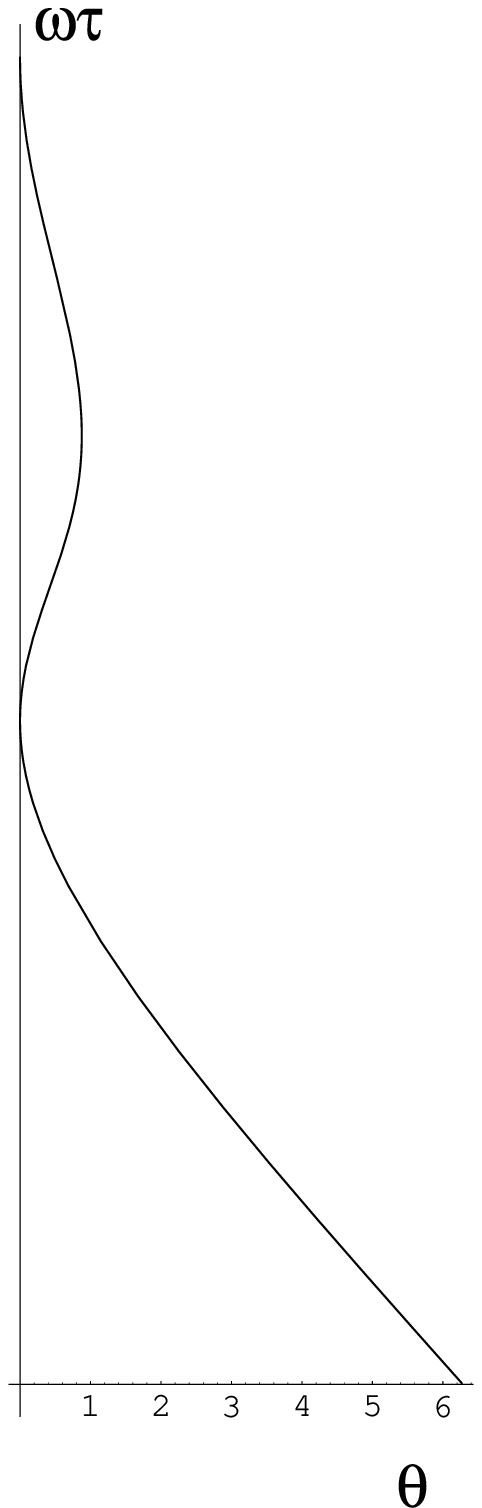,height=8cm}
}
} }
\FIG{fig:rotp1}{\figrotpi}{\FIGrotpi}

To have a solution that returns to $\r_1 \not = 0$ after time $\tau$
we must choose the initial momenta according to $\r_q = \r(\tau)$ and 
with (\ref{eqn:RotJac}) this gives
\[ \label{eqn:findp0}
        \r_1 = \D(\omega \tau) \r_1 + \tau \D(\omega \tau) \P_1.
\]
The final momenta are $\P(\tau) = \D(\omega \tau) \P_1$ and the angle 
$\theta$ between the initial and final velocities $\v=\P-\Ap(\r_1)$ is given by
\[
        \cos\theta = 1 - 2\frac{(1-\cos\omega\tau)^2}
        {2-2\cos\omega\tau - 2\omega\tau \sin\omega\tau + \omega^2\tau^2}.
\]
For $\omega\tau = 2\pi m$, $m=\pm1,\pm2,\ldots$, $\theta$ is zero such 
that the free motion is periodic by itself; in the non rotating frame
the particle does not move at all.
This is similar to the ordinary billiard,
where an orbit touching the boundary becomes infinitely unstable 
due to $\n \rightarrow 0$.
For every other value of $\tau$, however, a periodic orbit of
the billiard exists. 
The angle between the normal at the reflection point and the velocity
in the non rotating frame (!) is given by 
$\vartheta = (\pi-\omega\tau)/2 \bmod \pi - \pi/2$.
For negative $\vartheta$ the velocity is directed against the direction of
rotation of the disk (retrograde), while for positive $\vartheta$ the 
direction is the same (direct).
Retrograde orbits typically have higher energy than direct orbits
because they have to be faster to reach the next collision.
The energy for these periodic orbits parametrized by $\tau$ is
obtained by eliminating $\P$ in (\ref{eqn:RotHam}) using (\ref{eqn:findp0}),
\[ \label{eqn:hrot}
  h = \frac{\rr^2}{\tau^2}(1 - \cos\omega\tau - \omega\tau\sin\omega\tau),
\]
which is shown in \fig~\ref{fig:rotp1}, where $|\r_1| = \rr = 2$.
Every intersection of the line $h=\const$ with this curve corresponds to
a periodic orbit of the system.
At extrema of this curve periodic orbits are created/destroyed in pairs when
the energy is varied. At maxima retrograde orbits bifurcate and
at the minima direct orbits bifurcate.
These important points are given by 
\[
        h' = \frac{\partial h}{\partial \tau} = -\frac{\rr^2}{\tau^3}
(2-2\cos\omega\tau - 2\omega\tau\sin\omega\tau + \omega^2\tau^2\cos\omega\tau)
\stackrel{!}{=} 0,
\]
which will turn out to be closely related to the stability.

Taking the trace of $\K \M$ we find the general stability formula to be
\[ \label{eqn:Trotp1}
        \Trf_1 = \tau \left( 
                \cos\omega\tau (\CKe_{11} + \CKe_{22}) +
                \sin\omega\tau (\CKe_{21} - \CKe_{12}) \right),
\]
and for the use of (\ref{eqn:trCK}) we need
$\SP{\pr U}{\N} = \omega^2(\qrx_1\sin\alpha - \qry_1\cos\alpha)$ and
$\Bmag = -2\omega$.
In order for the reflection to map
$\v(\tau)$ into $\v(0)$ we must have 
$\SP{\P(\tau)-\Ap(\r_1)}{\T_1} = 
 \SP{(\P_1-\Ap(\r_1)}{\T_1}$,
which gives $\alpha_1=\arctan(\qrx_1/\qry_1)$. 
We now choose a coordinate system in such a way that
the orbit starts at $\qry_1=0$, such that $\alpha_1 = \pi/2$ and $\D = \S$.
Introducing the distance from the origin $r=\qrx_1$ the residue 
obtained from (\ref{eqn:Trotp1}) is
\begin{eqnarray} \label{eqn:rotp1}
\GR & = & \frac{1}{2} \frac{2 - 2\cos\omega\tau - 2\omega\tau\sin\omega\tau +
                        \omega^2\tau^2\cos\omega\tau}
                {\cos\omega\tau - 1}
        (r \kappa  - 1) \\
& = & (\omega^2\tau^2/2 - \frac{\cos\omega\tau-1}{\cos\theta-1})
                (r\kappa - 1) \\
& = & h' \frac{\tau^3}{2 \rr^2} \frac{r\kappa-1}{1-\cos\omega\tau}.
\end{eqnarray}
The last formula nicely illustrates the saddle-center bifurcations 
at the extrema of $h(\tau)$: Two new orbits are created
with stability $\GR=0$, i.e.\ $h'=0$, which is just the condition
for the bifurcation. Close to the bifurcation point $h'$ has different
signs for the two orbits, such that one is elliptic and the other one
direct hyperbolic. The elliptic orbit becomes inverse hyperbolic
typically after only a slight change of $\tau$. The interval of 
stability decreases with increasing $\tau$.
The whole scenario
is illustrated in the stability diagram of \fig~\ref{fig:rotp1}, which
completes the results reported in \cite{MBLTJS95}.

As our final application we now turn to the calculation of the 
period two orbits in the R3BB, which are of great
interest because of their apparent similarity with orbits in the
restricted three body problem that have close encounters with
both centers. We treat the symmetric case with 
$\kappa = \kappa_i$ and $\r_1 = (r,0)$ and $\r_2 = (-r,0)$.
The analogue of (\ref{eqn:findp0}) gives $\P_1$ for
the given $\r_1$ and $\r_2$. The resulting energy is
\[ \label{eqn:h2rot}
  \tilde h = \frac{\rr^2}{\tau^2}(1 + \cos\omega\tau + \omega\tau\sin\omega\tau).
\]
The main difference to the period one orbits is that $\tilde h$ diverges 
for $\tau \rightarrow 0$. Therefore we have a period two orbit up
to infinite energies, which bounces back and forth between the two
disks with the motion becoming a straight line for $\tau \rightarrow 0$
and $\tilde h \rightarrow \infty$. Except for this behavior we again have
the oscillations in $\tilde h(\tau)$ and therefore a 
bifurcation scenario similar to that of the period one orbits.
The important features of the residue again depend on the derivative 
of $\tilde h$, which is given by
\[
        \tilde h' = \frac{\partial \tilde h}{\partial \tau} = -\frac{\rr^2}{\tau^3}
(2+2\cos\omega\tau+2\omega\tau\sin\omega\tau-\omega^2\tau^2\cos\omega\tau).
\]
Because of the above symmetry assumptions the $\T$-$\N$-coordinates are
given by $\alpha_1 = \pi/2$ and $\alpha_2 = 3\pi/2$ such that
$\K_1 = \K_2$ with $\W_1=\W_2=\W_0$. Although the orbit is not
time reversal symmetric we have $\tau = \tau_1 = \tau_2$ due to the
spatial symmetry and therefore $\M_1=\M_2$. 
For the trace of $\K\M\K\M$ we find
\[
  \Trf_2 = 4 (2\bar\GR'-1)^2
, \quad
\bar\GR' = \tilde h' \frac{\tau^3}{2 \rr^2} \frac{r\kappa-1}{1+\cos\omega\tau},
\]
giving the same factorization of $\bar\GR=\Trf_2/4$ as 
in the gravitational billiard (\ref{eqn:factori}). 

The bifurcation scenario for the period two orbits is a follows. They are
created in saddle center bifurcations at the extrema of $\tilde h$. 
The residue 
of the elliptic orbit increases from 0 to 1. However the residue does not
increase beyond 1 but instead begins to decrease again. This 
``touching of residue 1'' is an effect of the symmetry of the orbit,
which corresponds to passage of $\GR'$ through $1/2$. 
There are two period four orbits involved in this $2/4$-bifurcation
\cite{GMVF81} and the old orbit stays elliptic. Eventually it reaches
$\GR=0$ again and looses stability in a period doubling bifurcation.
For retrograde orbits this scenario happens for decreasing period $\tau$, 
and for direct orbits with increasing period.
Decreasing the period of the retrograde orbits further eventually turns
them infinitely unstable at $1+\cos\omega\tau=0$, where they have a
tangency instead of a reflection. The same happens for the direct
orbits if we increase their period. However, if we increase the period
of a retrograde orbit it becomes more and more unstable, but eventually
becomes direct and then stable again, thus every two of the
orbits described above are continuously connected in phase space. 
The whole scenario is repeated 
for any number $m$ of revolutions that are completed in the non rotating
frame before the next reflection. The intervals of stability are
extremely small for higher periods. Only in the case $\kappa r \approx 1$,
i.e.\ the distance from the origin is close to the radius of curvature,
are the windows of stability larger. For $\kappa r = 1$ all the
orbits become inverse parabolic, independently of the period. Note that this
case is completely different from the standard case in the R3BB, 
because the curvature must be positive.
As a matter of fact, the case $\kappa r =1$ can be realized as an
orbit {\em inside} a circle, and we reobtain the integrable circle 
billiard described in rotating coordinates.
The distance between the reflection points of the orbit is $l=2r$,
such that the same geometric factor enters the residue as in the 
previous examples. 

It is possible to treat the general period two orbit, with
different curvatures and distances from the origin. This case is
quite promising in order to make a more quantitative comparison
with the restricted three body problem, because with these parameters
we have an analog of the mass ratio in our system.
In any case the system at hand, which was introduced in \cite{MBLTJS95},
gives the opportunity to obtain analytical stability results.
This tempers the feeling that this approximation of the three body problem
is too crude, and we hope, as cited in the beginning, that ``only the 
interesting qualitative questions need to be considered''.

\section{Conclusion}

We have derived a general expression for the linearized Poincar\'e map
in a billiard with potential. Due to the additional coupling introduced
by a (vector) potential the linearized map can only conveniently be written
in four dimensions. Using a canonical transformation
in phase space we have calculated the linearization and it turned out
that the arc length and the parallel component 
of the velocity (not of the momentum) give a canonical coordinate 
system on the surface of section given by the billiard boundary.
This holds true for any potential. The resulting formula is useful
for the application of numerical methods to find periodic orbits,
or for the estimation of the Lyapunov exponent. Moreover, since
the linearization in phase space includes the variation of the energy, 
which usually gives a nontrivial change in the orbit for a billiard with 
potential, it can also be used to follow periodic orbits when the energy 
is varied.

The main application of the linearized map is the calculation of the
stability of periodic orbits. The trace of the (four dimensional) monodromy 
matrix of the periodic orbit can be factorized 
into a product of matrices describing the piecewise free
motion between reflections and the contribution from the reflection.
This generalizes a well known formula with $2\times 2$ matrices for 
ordinary billiards. 
With these results it was possible to perform the stability analysis 
of period one and two orbits in four billiards with 
different (vector) potentials. 

The results are presented in form of stability diagrams, where the
residue of the orbits is shown in its essential parameter dependence.
These diagrams can give some intuition about the 
mechanisms that create stability respectively instability.
The investigation of stability diagrams of the ordinary billiard
(which is the same as for the one with magnetic field) and
the gravitational billiard showed that periodic orbits in one parameter 
families of integrable systems tend to avoid period doubling bifurcations 
which are generically present.
In the final application to a rotating billiard, which mimics the
restricted three body problem, we have obtained an interesting 
bifurcation scenario of an infinite number of period one and two orbits
by successive saddle-center bifurcations with increasing period.
The calculations for this system can be extended to the non-symmetric case, 
in order to obtain a simple system that presents some 
essential features of the restricted three body problem in an
analytically tractable way.

\ack

The author would like to thank 
A. Wittek,
P.H. Richter,
H.-P. Schwebler,
and
A. B\"acker
for helpful discussions.
This work was supported by the Deutsche Forschungsgemeinschaft.

\secref

\Capts{ 
\newpage
\section*{Figure captions}
\setlength{\parindent}{0cm}
\printfigcap{fig:notation}{\fignotation}
\printfigcap{fig:bilp2}{\figbilpii}
\printfigcap{fig:grap2}{\figgrapii}
\printfigcap{fig:rotp1}{\figrotpi}
\clearpage\newpage
\pagestyle{empty}
\showfig{\FIGnotation}
\showfig{\FIGbilpii}
\showfig{\FIGgrapii}
\showfig{\FIGrotpi}
}

\end{document}